# BACOM2: a Java tool for detecting normal cell contamination of copy number in heterogeneous tumor

Yi Fu[1]†, Jun Ruan[2]†, Guoqiang Yu[1], Douglas A. Levine[3], Niya Wang[1], Ie-Ming Shih[4], Zhen Zhang[4], Robert Clarke[5], and Yue Wang[1]

[1]Department of Electrical and Computer Engineering, Virginia Polytechnic Institute and State University, Arlington, VA 22203, USA; [2]School of Information Engineering, Wuhan University of Technology, Wuhan, 430070; [3]Department of Surgery, Memorial Sloan-Kettering Cancer Center, New York, NY 10021, USA; [4]Departments of Pathology and Oncology, Johns Hopkins University, Baltimore, MD 21231, USA; [5]Lombardi Comprehensive Cancer Center, Georgetown University, Washington, DC 20057, USA;

## Abstract

**Background**

BACOM is a statistically principled and unsupervised method to estimate genomic deletion type and normal tissue contamination, and accordingly recover the true copy number profile in cancer cells [1]. However, the average normal cell fraction estimated by BACOM was found higher than expected in TCGA ovarian cancer dataset.

**Results**

We develop a cross-platform open-source Java application (BACOM2) with graphic user interface (GUI), and users also can use a XML file to set the parameters of algorithm model, file paths and the dataset of paired samples. BACOM2 implements the new entire pipeline of copy number change analysis for heterogeneous cancer tissues, including extraction of raw copy number signals from CEL files of paired samples, attenuation correction, identification of balanced AB-genotype loci, copy number detection and segmentation, global baseline calculation and absolute normalization, differentiation of deletion types, estimation of the normal tissue fraction and correction of normal tissue contamination.

**Conclusion**

BACOM2 focuses on the common tools for data preparation and absolute normalization for copy number analysis of heterogeneous cancer tissues. The software provides an additional choice for scientists who require a user-friendly, high-speed processing, cross-platform computing environment for large copy number data analysis.

## Background

Changes in the number of copies of genomic DNA is an important step in the progression of cancer. Quantitative analysis of somatic copy number alterations (CNAs) has found broad application in cancer research. However, tissue samples often consist of mixed cancer and normal cells, and accordingly, the observed SNP intensity signals are the weighted sum of the copy numbers contributed from both cancer and normal cells. The intertwined result is that copy number aberrations, and tumor aneuploidy, the average ploidy of tumor cells cannot be assumed to be 2N or integer [2].

Yu et al. proposed a statistically principled in silico approach, Bayesian Analysis of COpy number Mixtures (BACOM), to accurately estimate genomic deletion type and normal tissue contamination, and accordingly recover the true copy number profile in cancer cells [1]. However, the normalization baseline of copy number of all chromosomes selected the mean of the biggest component of the copy number intensities in BACOM, and the assumption is that the majority of loci are normal. As a result, the average normal cell fraction estimated by BACOM was found higher than expected in TCGA ovarian cancer dataset when the biggest copy region did not correspond to normal but triple or fourfold. Thus, we report here an improved method to optimize normalization baseline and parameter estimation, and describe an allele-specific absolute normalization and quantification scheme that can enhance BACOM applications in many biological contexts. We develop a cross-platform open-source Java application (BACOM2) that implements the evolutionary pipeline of copy number analysis of heterogeneous cancer tissues including relevant processing steps.

## Implementation

The open-source BACOM2 application was written in the Java programming environment. The BACOM2 allows the user to use GUI (Figure 1) to run the application, and also can use a XML file to set the parameters of algorithm model, file paths and the dataset of paired samples. The BACOM2 software implements the new entire pipeline of copy number change analysis for heterogeneous cancer tissues, including extraction of raw copy number signals from CEL files, attenuation correction, identification of balanced AB loci, copy number detection and segmentation, probe sets annotation, global baseline calculation and normalization, differentiation of deletion types, estimation of the normal tissue fraction and correction of normal tissue contamination. Interested readers can freely download the software and source code at: https://code.google.com/p/bacom2/.

The different of overall framework to original BACOM, BACOM2 only uses parallel computing to segment each chromosome, and other processes are executed in the order of sequence for helping to reduce disk I/O. When handing multiple paired samples, BACOM2's processing speed increases up to three times. Another feature of BACOM 2 is complete logging and multiple output charts, including histogram of copy number, LRR and BAF chart of Copy number profile, etc.

### Method overview

Tissue samples often consist of mixed cancer and normal cells, and accordingly, the observed SNP intensity signals are the weighted sum of the copy numbers contributed from both cancer and normal cells.

$$X_i = \alpha \times X_{normal,i} + (1 - \alpha) \times X_{cancer,i} \qquad (1)$$

Where $X_i$ is the observed copy number signal at locus $i$, $\alpha$ is the unknown fraction of normal cells, $X_{normal,i}$ and $X_{cancer,i}$ are the latent copy number signals in normal and cancer cells at locus $i$, respectively. Let $X_{A,i}$ and $X_{B,i}$ be the signals of alleles A and B at locus $i$, then $X_i$ is the sum of $X_{A,i}$ and $X_{B,i}$ at each locus $i$. A chromosome is generally divided into a few continuous segments using copy number data, within each of which the copy numbers are considered constant. In each segment, assume $X_{A,i}$ and $X_{B,i}$ are Gaussian random variables with distinct means but common variance, and $X_i$ also are Gaussian random variable following a normal distribution N($\mu_{A+B}, \sigma^2_{A+B}$) whose mean and variance can be readily estimated.

When all chromosomes of a sample are encapsulated into some segments, we calculate Pearson's correlation coefficient ρ between $X_{A,i}$ and $X_{B,i}$ in the range of a segment. We can filter out the allelic-balanced loci by whether ρ sufficiently close to 1. The allelic-balanced loci means two alleles A and B at the same locus on homologous chromosomes are expressed at the same level, for example, AB, AABB(duplication), AAABBB(triploid),--(loss). Tested on many real copy number datasets, we found that the dominant component of the histogram model of the genome-wide copy number signals of the allelic-balanced loci generally corresponds to the normal copy number or the fourfold-higher copy number regions in majority of cancer types. We filter out the two largest peaks of copy number signals histogram of the allelic-balanced loci, then the lower copy number value of the two peaks correspond to normal (CN = 2.0); the higher copy number roughly correspond to duplication (CN = 4.0). If there is only one peak, the copy number correspond to it will be judged by the position of the histogram of all loci. And we adopt a relatively loose range of robust means to determine the normal segments, that is, $\pm 15\%$.

When the segments with robust means sufficiently less than normal, the segments can be determined as deletion segments. Allele-specific analyses are focused on the allelic-balanced loci of the deletion regions. Types of deletions are detected by a model-based Bayesian hypothesis testing. Specifically, BACOM uses a novel summary statistic random variable $Y, Y > 0$,

$$Y = \sigma_{A-B}^{-2} \sum_{i=1}^{L} (X_{A,i} - X_{B,i})^2 = \frac{1+\rho}{1-\rho} \sigma_{A+B}^{-2} \sum_{i=1}^{L} (X_{A,i} - X_{B,i})^2 \qquad (2)$$

where $\sigma_{A-B}^2$ is the variance of $X_{A,i} - X_{B,i}$ in a length-$L$ deletion region, $\sigma_{A+B}^2$ is the variance of $X_{A,i} + X_{B,i}$, ρ is the correlation coefficient between $X_{A,i}$ and $X_{B,i}$. The summary statistics Y within a Length-$L$ hemi-deletion segment follows an L degrees of freedom non-central $\chi^2$ distribution, given by (here y > 0):

$$\chi^2(y; L, \lambda) = \frac{e^{-(y+\lambda)/2}}{2^{L/2}} \sum_{k=0}^{\infty} \frac{y^{\frac{L}{2}+k-1} \lambda^k}{\Gamma(k + L/2) 2^k k!} \qquad (3)$$

where $\lambda = L(2 - \mu_{A+B})^2 \sigma_{A+B}^{-2} (1 + \rho)/(1 - \rho)$, ρ is the correlation coefficient between $X_{A,i}$ and $X_{B,i}$, and Γ denotes the Gamma function. And Y within a Length-L homo-deletion segment follows an L degrees of freedom standard $\chi^2$ distribution, given by (here y > 0):

$$\chi^2(y; L) = \frac{1}{2^{L/2} \Gamma(L/2)} y^{\frac{L}{2}-1} e^{-\frac{y}{2}} \qquad (4)$$

For a deletion region, using the two $\chi^2$ distribution equations we can respectively calculate two posterior probabilities of a deletion segment with its summary statistic random variable Y. Then, a straightforward application of Bayesian hypothesis testing implies the deletion type of the segment.

Since for a deletion region, we have:

$$\begin{cases} \mu_i = \alpha \times 2 + (1-\alpha) \times 0 = 2\alpha, & if\ homo-deletion, \\ \mu_i = \alpha \times 2 + (1-\alpha) \times 1 = 1+\alpha, & if\ hemi-deletion, \\ \mu_i = \alpha \times 2 + (1-\alpha) \times 4 = 4-2\alpha, & if\ duplication, \end{cases} \quad (5)$$

where $\mu_i$ is the sample average of the copy number signals of the $i$-th length-$L$ deletion segment, $\mu_i = (1/L)\sum_{i=1}^{L} X_i$. Subsequently, we can estimate the fraction of normal cells of each deletion segments in the sample by (5); and we can calculate the fraction of normal cells of each duplication segments which is nearby the higher peak of copy number histogram of allelic-balanced loci. Then, the average normal cell fraction $\bar{\alpha}$ across the whole genome is the median of all calculated α. Finally, the true copy numbers of pure cancer cells can be recovered by $\widehat{X}_{cancer,i} = (X_i - 2\bar{\alpha})/(1-\bar{\alpha})$.

The new entire pipeline of BACOM2 is summarized in Figure 2.

**Data preprocessing**

BACOM 2 uses allelic information (allelic summary statistics, allele-balance/imbalance, inter-allele correlation) to determine deletion type and subsequent analyses. Reading raw data from '.CEL' files in BACOM 2 is very similar to original BACOM [1, 3] except using different file format of the platform annotation library provided by Affymetrix, which is used to extract the optical scanned intensity value for each probe. Probes' chromosome IDs, locations and probe set information. Through correcting positive noise offset and identifying SNP probe sets with genotype AB for the raw observed signal intensity, we use the Langmuir isothermal absorption model to correct for signal attenuation [4, 5].

In BACOM 2, absolute signal normalization is performed based on the identified signal mean of individual copy-neutral segments (corresponding to true copy number '2') [1, 6]. The different from original BACOM, here we perform signal segmentation before signal normalization [7], replace original recursive process by an iterative algorithm, adopt parallel computing each chromosome, add a merge operation after original segmentation and optimize the original segmentation algorithm of BACOM by using an integral array to accelerate summations.

Here we can use the parameter θ to control the result of segmentation, as shown in Figure 3. The parameter θ is used to constrain the minimum difference of copy number between two adjacent segments. When the difference of copy number between two adjacent segments is too low, then we merge these two segments into one segment. In BACOM 2, the count of segments of all chromosomes has limited impact on the subsequent processing, as long as the length of each segment is not too short. Generally the range of θ is from 0.5 to 0.8 by conservative principles.

**Normalization**

To perform accurate absolute normalization, identification of copy-neutral or normal copy number loci is a critical step. We now know that the dominant component of the intensity mixture distribution rarely coincides with the normal copy number '2' [2]. Our experimental studies on real tumor data confirmed this observation, while also indicated that the largest component(s) often resides within the neighbourhood of normal copy number component. In BACOM 2, we performed three steps to correctly filter out potential copy-neutral component from all SNP Loci, as shown in Figure 4.

Our method focus solely on AB-genotype probe sets, firstly all AB-genotype loci of normal sample are labeled by a genotype calling algorithm, which cluster $\log_2$ of the raw allele intensity data into three groups, AA-genotype, BB-genotype and AB-genotype, by distance from data point to diagonal. Secondly, we calculate allele intensity signals of paired sample, and the allele-balanced loci are selected by calculating Pearson's correlation coefficient ρ between $X_{A,i}$ and $X_{B,i}$ in the range of a segment. We can filter out the allelic-balanced loci by whether ρ sufficiently close to 1, through selecting an appropriate sliding windows width for calculating ρ. We generally select the rightmost peak in the histogram of ρ as the allele-balanced part, as shown in Figure 5(a). This step will largely remove majority of allele-imbalanced loci including copy-neutral LOH. Finally, with the histogram of copy number of allele-balanced loci, the true copy-neutral component is highly probably to be the most dominant component or the second most dominant component. In fact, among the components associated with allele-balanced loci, homo-deletion segments correspond to '0' and sixflod segments (AAABBB-genotype) correspond to '6' are usually rare and short that can be confirmed by examining real copy number datasets. So our main concern is the observed copy number correspond to '2' and '4'.

Here if two peaks can be detected, we can correctly identify the true copy number of copy-neutral component by three numerical difference of zero and two peaks. As shown in Figure 5(b), the most dominant component of allele-balanced loci correspond to copy number '4', and the second most dominant component correspond to copy number '2'.

If only one peak is detected, we will find out the corresponding position of possible copy-neutral component in the histogram of copy number of all SNP loci. If here are more than two peaks is less than the copy number of candidate "copy-neutral" component, the candidate "copy-neutral" component should correspond to copy number '4' instead of copy number '2'.

When we identify the copy number component corresponding to '2', a genome-wide absolute normalization will be performed by local baseline of each chromosome. And the deletion segments will be determined naturally by specified range.

### Normal tissue fraction estimation

We exploited a mathematically-justified scheme to correct for the confounding impact of intratumor heterogeneity on estimating tumor purity [2, 8]. Slight different from original BACOM, apart from Homo-deletion segments and Hemi-deletion segments, we also introduce duplication segments to estimate normal fraction α. Thus, when deletion segments do not exist, the application can still calculate normal fraction α by duplication segments. Based on the distribution of α estimates across the whole genome, BACOM 2 calculate the final value of normal fraction using the median of α estimates. The final recovered result is shown in Figure 5(c).

## Results and Discussion

All the real datasets used in this section are from The Cancer Genome Atlas (TCGA) database, acquired by Affymetrix Genome-Wide Human SNP Array 6.0. In BACOM 2 analysis, we use only matched tumor-normal pairs since somatic alterations are of the interest.

We analysed total 466 samples in TCGA_OV dataset. And we compared the results obtained by BACOM2 with the previously reported results obtained by ABSOLUTE (Carter, et al., 2012). Tested on the same tumor sample as shown in Figure 6.

With a quality control selection on paired tumor and normal samples, ABSOLUTE analysed 410 tumor samples in the OV dataset. The average tumor purity estimates by BACOM 2.0 and ABSOLUTE are 51% and 78%, respectively; and the average tumor ploidy estimates by BACOM 2.0 and ABSOLUTE are 2.61 and 2.77, respectively. The sample-wise correlation coefficients show that both tumor purity and tumor ploidy estimates by BACOM 2.0 correlate well with the estimates by ABSOLUTE, achieving a high correlation coefficient of $r = 0.67$ on purity and a high correlation coefficient of $r = 0.77$ on ploidy. We argue that the estimated bias of purity is caused by using completely different data preprocessing methods, specifically attenuation correction.

There are some questions worth further exploration. For determining the copy-neutral component, all probe set signals were filtered three times according to genotype, allelic-balanced level and frequency. On the one hand, the computational efficiency is relatively very high along with each filter to reduce the amount of loci that need to be calculated in next step. On the other hand, if the effective allelic-balanced loci are too little, it cause two largest component of histogram of allelic-balanced loci is no longer reliable due to noise. Even allelic-balanced loci are absence in a very few paired samples. We expect an improved filtering steps to see further developments. Specifically, the usage of extreme imbalanced loci is issues that remain open.

## Conclusions

We develop a cross-platform open-source Java application (BACOM2) with graphic user interface (GUI) to estimate copy number deletion types and normal tissue contamination, and to extract the true copy number profile in cancer cells. BACOM2 implements the new entire pipeline of copy number change analysis for heterogeneous cancer tissues, including extraction of raw copy number signals from CEL files of paired samples, attenuation correction, identification of balanced AB-genotype loci, copy number detection and segmentation, global baseline calculation and absolute normalization, differentiation of deletion types, estimation of the normal tissue fraction and correction of normal tissue contamination. It is easy to use and requires no prior specific computer science knowledge. A user- friendly interface simplify parameter settings and management of raw data. Parallel processing in BACOM 2 allows for the analysis of multiple chromosomes in one paired sample at the same time. BACOM2 multiple output charts and complete logging. We expect that, with further development, BACOM2.0 software tool would find broad biomedical applications [9].

**Funding**: National Institutes of Health, under Grants CA149147, CA160036, HL111362, NS29525, in part.

**Conflict of Interest**: none declared.

## References

[1]  G. Yu, B. Zhang, G. S. Bova, J. Xu, M. Shih Ie, and Y. Wang, "BACOM: in silico detection of genomic deletion types and correction of normal cell contamination in copy number data," *Bioinformatics,* vol. 27, pp. 1473-80, Jun 1 2011.

[2]  M. Rasmussen, M. Sundstrom, H. Goransson Kultima, J. Botling, P. Micke, H. Birgisson, B. Glimelius, and A. Isaksson, "Allele-specific copy number analysis of tumor samples with aneuploidy and tumor heterogeneity," *Genome Biol,* vol. 12, p. R108, 2011.

[3]  N. Wang, T. Gong, R. Clarke, L. Chen, I. M. Shih, Z. Zhang, D. A. Levine, J. Xuan, and Y. Wang, "UNDO: a Bioconductor R package for unsupervised deconvolution of mixed gene expressions in tumor samples," *Bioinformatics,* p. DOI 10.1093/bioinformatics/btu607, Sep 10 2014.

[4]  S. L. Carter, K. Cibulskis, E. Helman, A. McKenna, H. Shen, T. Zack, P. W. Laird, R. C. Onofrio, W. Winckler, B. A. Weir, R. Beroukhim, D. Pellman, D. A. Levine, E. S. Lander, M. Meyerson, and G. Getz, "Absolute quantification of somatic DNA alterations in human cancer," *Nat Biotechnol,* vol. 30, pp. 413-21, May 2012.

[5]  D. Hekstra, A. R. Taussig, M. Magnasco, and F. Naef, "Absolute mRNA concentrations from sequence-specific calibration of oligonucleotide arrays," *Nucleic Acids Res,* vol. 31, pp. 1962-8, Apr 1 2003.

[6]  B. Zhang, X. Hou, X. Yuan, M. Shih Ie, Z. Zhang, R. Clarke, R. R. Wang, Y. Fu, S. Madhavan, Y. Wang, and G. Yu, "AISAIC: a software suite for accurate identification of significant aberrations in cancers," *Bioinformatics,* vol. 30, pp. 431-3, Feb 1 2014.

[7]  X. Yuan, G. Yu, X. Hou, M. Shih Ie, R. Clarke, J. Zhang, E. P. Hoffman, R. R. Wang, Z. Zhang, and Y. Wang, "Genome-wide identification of significant aberrations in cancer genome," *BMC Genomics,* vol. 13, p. 342, 2012.

[8]  L. Oesper, A. Mahmoody, and B. J. Raphael, "THetA: Inferring intra-tumor heterogeneity from high-throughput DNA sequencing data," *Genome Biol,* vol. 14, p. R80, Jul 29 2013.

[9]  Y. Tian, S. Wang, Z. Zhang, O. Rodriguez, E. Petricoin III, I.-M. Shih, D. Chan, M. Avantaggiati, G. Yu, S. Ye, R. Clarke, C. Wang, B. Zhang, Y. Wang, and C. Albanese, "Integration of network biology and imaging to study cancer phenotypes and responses," *IEEE/ACM Transactions on Computational Biology and Bioinformatics,* vol. 11, pp. 1009-1019, 2014.

# Figures

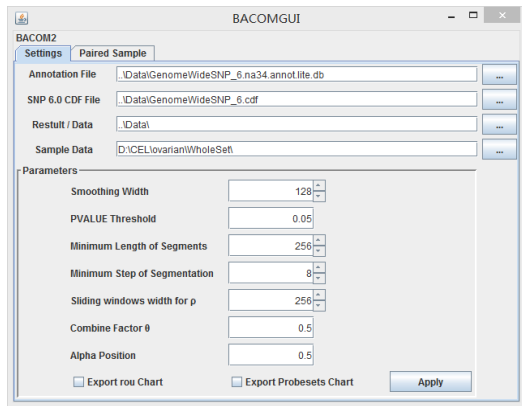

**Figure 1** GUI of BACOM2

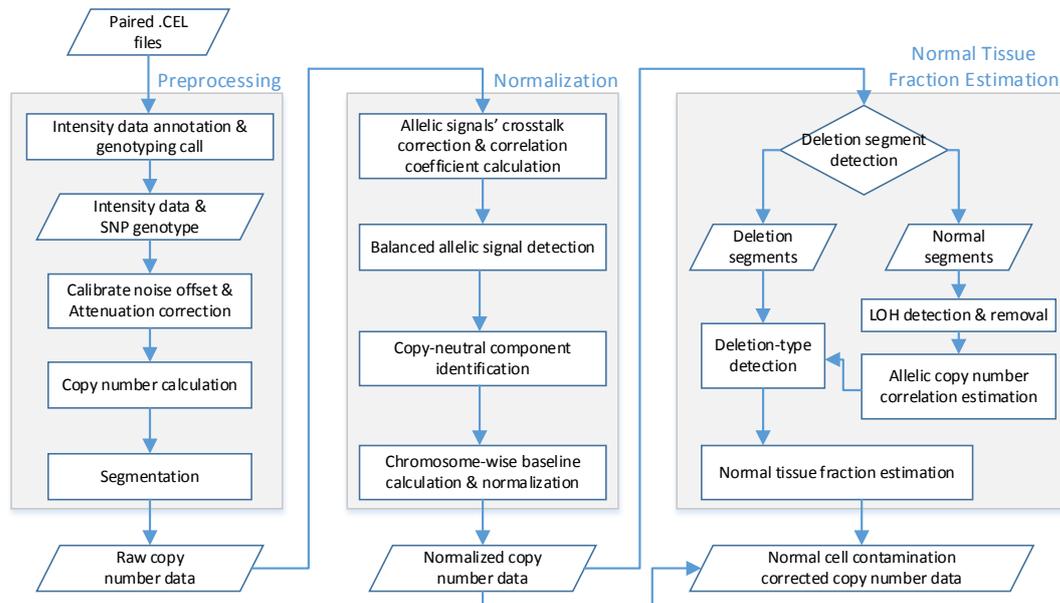

**Figure 2** Analytic pipeline of BACOM2: schematic flowchart

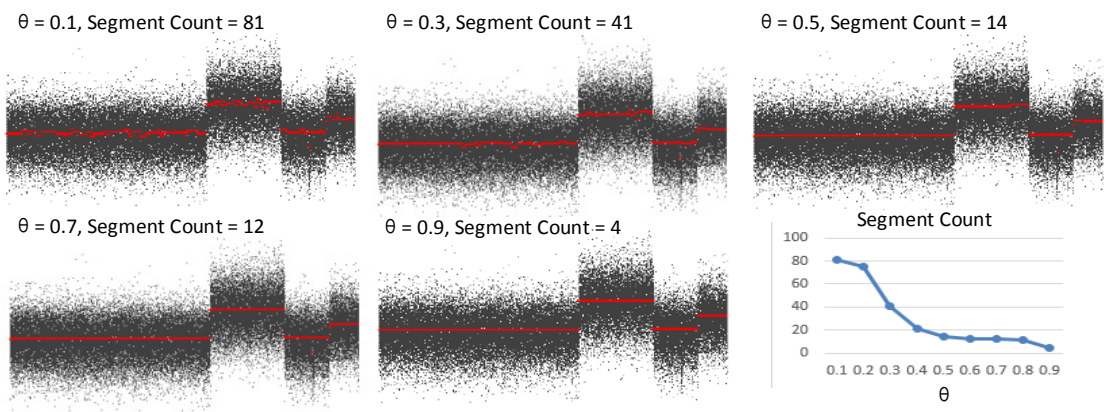

**Figure 3** The effect of changing the parameter θ in segmentation algorithm.

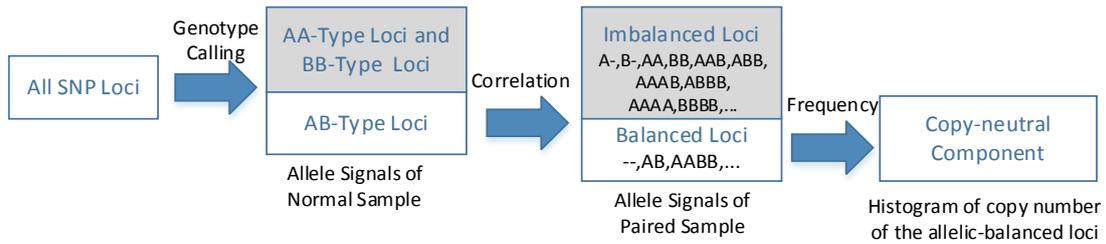

**Figure 4** Extract copy-neutral component by filtering allele-imbalanced signals.

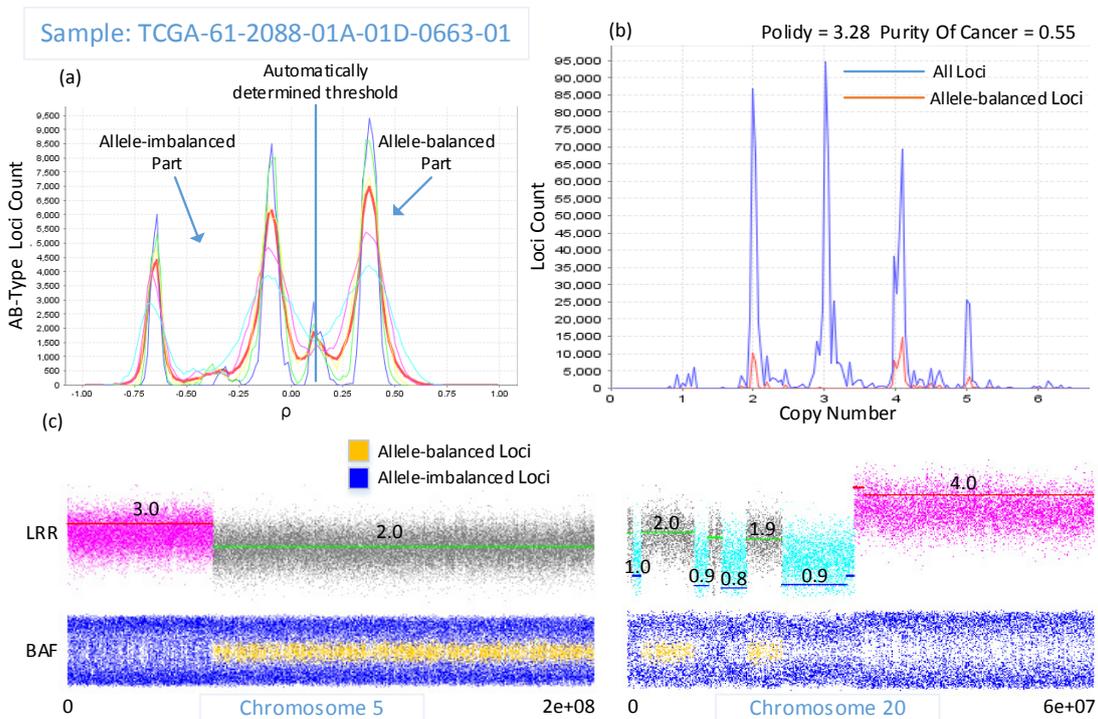

**Figure 5** (a) Histogram of correlation coefficient ρ of AB-genotype Loci at different window sizes, the red thick line is average value of ρ frequency at five different window sizes (64, 128, 256, 512, 1024); (b) Histogram of copy number after correction by cancer purity; (c) Copy number profile and correction results.

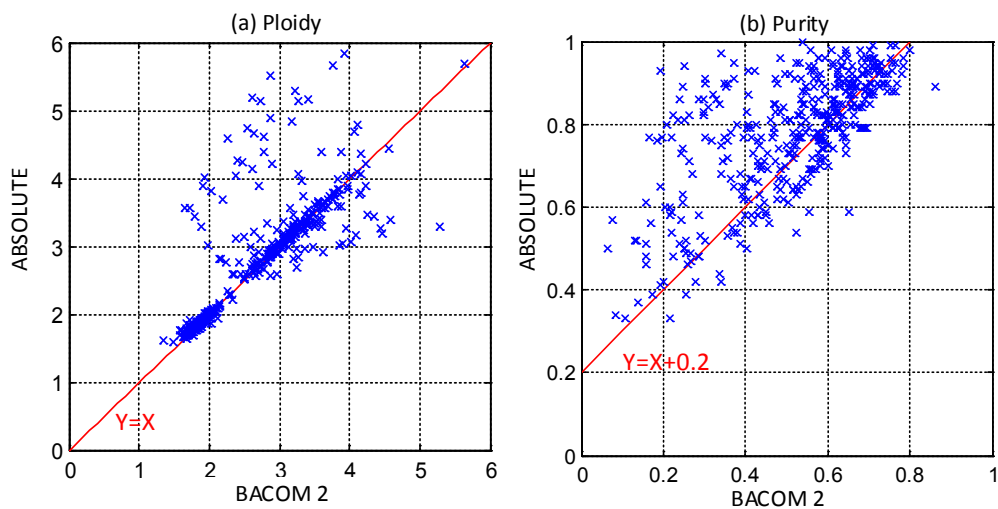

**Figure 6** Sample-wise comparison between the estimates of tumor purity and average ploidy by BACOM 2 and ABSOLUTE on TCGA ovarian cancer samples. (a) Scatter plot of tumor purity estimates; (b) Scatter plot of tumor ploidy estimates.